# Cascaded multi-harmonic generation in a β-BBO crystal reaching 133 nm from a single 800 nm pump beam

J. Seres, E. Seres, T. Schumm

Atominstitut - E141, Technische Universität Wien, Stadionallee 2, 1020 Vienna, Austria

jozsef.seres@tuwien.ac.at

Nonlinear crystals are widely used to efficiently generate second or third harmonics, and also for frequency mixing. Generation of multiple cascaded harmonics simultaneously, however, appears much less studied, although supported by the usually large nonlinearity of these crystals. Here, we demonstrate the generation of multiple harmonics up to 6$^{th}$ order in a β-BBO crystal, extending into the vacuum ultraviolet spectral range. Phase matched generation of the second or the third harmonic is directly driven by a single 800 nm Ti:sapphire oscillator beam. Our study reveals that the harmonics are generated by $\chi^{(2)}{:}\chi^{(2)}$ and $\chi^{(2)}{:}\chi^{(3)}$ cascaded processes and intense harmonic signals are obtained even for the 5$^{th}$ and 6$^{th}$ harmonics, where the β-BBO crystal is mostly absorbing. While higher order harmonics cannot be phase matched simultaneously, limiting the harmonic amplitudes, even the weakest harmonics are more than three orders of magnitude above the measurement noise limit, making them suitable for many spectroscopic applications.

**Introduction**

In recent years, the use of solid materials for generating multiple and high order harmonics has attracted increased attention, offering an alternative to the widely studied gas-phase media. Different isolator [1] and semiconductor [2] materials, their thin films [3], nanomembranes [4], nanostructures, and even atomic layers were successfully tested. Beyond second harmonic generation, direct [5, 6, 7, 8] or cascaded third harmonic generation [9, 10] or four-wave mixing were examined to convert infrared laser beams into the ultraviolet spectral range, mainly for spectroscopic applications [11].

Multiple harmonic generation was theoretically predicted for up to 6$^{th}$ harmonic at 134 nm wavelength by non-collinear four-wave mixing in LiF [12], a material providing a large band gap and transparency at that wavelength. However, because of the relatively low damage threshold and cubic crystal structure, its direct multi-harmonics generation capacity is limited [3]. Transparent noble gas media were also attractive candidates for four-wave mixing [13, 14, 15] targeting vacuum ultraviolet (VUV) wavelengths. Furthermore, non-collinear phase matching for 3$^{rd}$ harmonic generation was demonstrated [16, 17]. For reaching higher harmonic orders, multiple nonlinear crystals were used subsequently, each of them cut and adjusted separately for phase matching [18, 19, 20, 21, 22, 23] and up to 5$^{th}$ harmonic [24] and the VUV spectral range were reached.

Somewhat less attention was paid on using established nonlinear crystals for generating multiple harmonics. They are usually used for the generation of second harmonics or sum- or difference frequencies, but their capabilities for generation of multiple harmonics are rarely explored [25, 26]. An attractive candidate is the β-BBO crystal because of its broad transparency range, relatively large damage threshold and especially, because its large birefringence provides phase matched second and even third harmonic generation down to the ultraviolet spectral range. The crystal is extensively studied and the possible ways to generate third harmonics directly by third order or cascaded second order interactions were described in details [7, 27, 28, 29].

In this study, we examine the multiple harmonic generation capacity of β-BBO crystals, used extensively for conversion of femtosecond laser pulses by second and third harmonic generation. As mentioned above, up to 6$^{th}$ harmonics were theoretically predicted [12] in LiF crystals, using multiple input beams and non-collinear arrangement of pump beams, requiring a complex optical setup. In our study, we use only one input laser beam and, combined with the internally generated second harmonic beam, even and odd harmonics up to 6$^{th}$ order are generated by means of different $\chi^{(2)}:\chi^{(2)}$, $\chi^{(2)}:\chi^{(3)}$ cascaded nonlinear processes; even the $\chi^{(3)}:\chi^{(3)}$ cascade has been observed. After describing the experimental setup, the generation of harmonics is demonstrated and assigned to the generating cascade processes; the assignment is then corroborated by further measurements. We demonstrate that the presence of the strong second harmonic beam is essential for the efficient generation of multiple harmonics.

**Experimental setup**

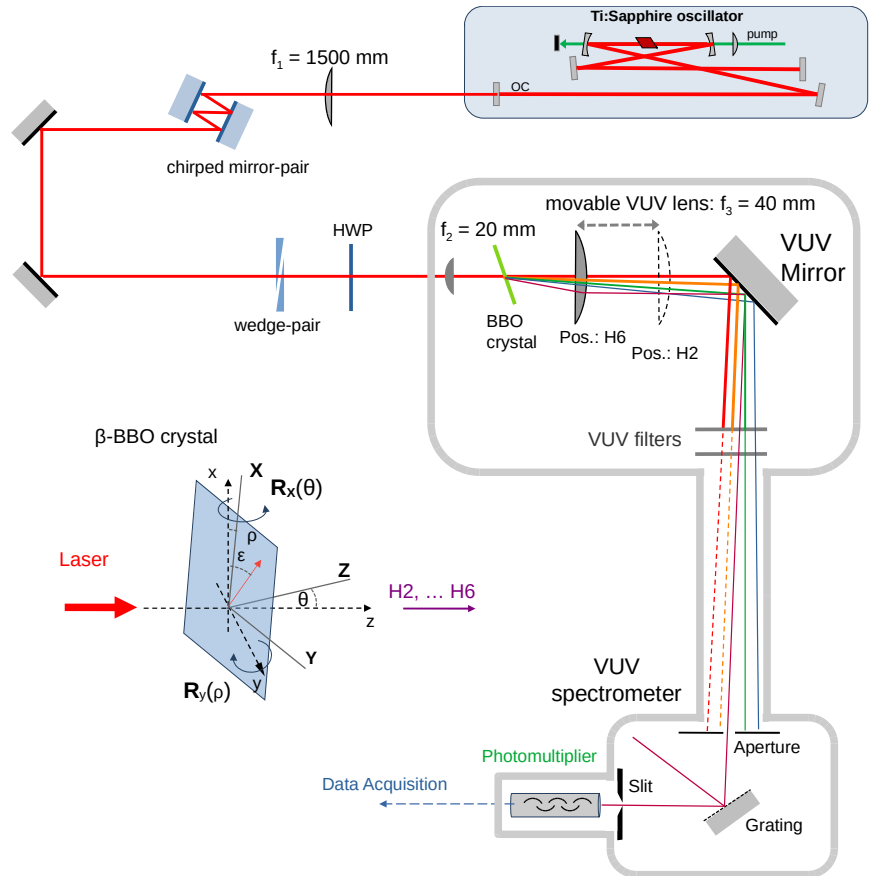

**Figure 1:** Experimental setup for the measurement of the multi-harmonic generation in a β-BBO crystal. HWP: half-wave plate. Inset: the nonlinear crystal with the notations used for describing the directions and rotations. H2, H3, … H6 denote the generated different harmonic orders.

In our experiments, a Ti:sapphire oscillator output of a frequency comb (Menlo Systems), running at 108 MHz repetition rate, was used. The output pulses were negatively pre-chirped using chirp mirrors and a BK7 wedge pair to precisely compensate the material dispersion of the different optical components (lenses, window), to obtain a compressed short pulse on the β-BBO non-linear crystal (Edmund Optics #23-340) placed into the focus. To the β-BBO crystal, 8 nJ, 28 fs ultrashort pulses at 800 nm central wavelength were delivered for harmonic generation. A small and thin lens with focal length of 20 mm focused the beam to ~10 µm (FWHM) spot size. The laser intensity in the focus was estimated approaching 200 GW/cm$^2$. The experimental setup is presented in Figure 1. Since we were interested in measuring the signals of all generated harmonics up to the vacuum ultraviolet (VUV) spectral range, parts of the experimental setup were placed into a vacuum chamber.

A VUV grade MgF$_2$ lens focused the generated harmonic beams onto an aperture (1 mm diameter), which served as the input to the spectrometer. By moving the lenses and turning the VUV mirror after the lens, it was possible to choose and focus the desired harmonic order to the input aperture, while the pump laser beam and the other harmonic beams became divergent and were spatially filtered out for higher signal/background contrast. The lenses and the VUV mirror in the vacuum chamber were placed onto motorized translation stages and mirror mount. Using additional VUV filters with a peak transmission wavelength of 130 nm (130-BB-1D, Pelham Research Optical L.L.C.), it was possible to further improve the measurement of the higher order harmonics by decreasing the contribution of the dominant second and third orders. The spectrometer contained a grating with 300 g/mm and an output slit before the Hamamatsu R6836 photomultiplier, which is mainly sensitive in the spectral range between 110 and 320 nm; for recording spectra, the grating was rotated. The combination of the solar-blind photomultiplier, spectral and spatial filtering allowed us to measure signal amplitudes over more than 10-orders of magnitude, from sub-W to the 10 pW levels. Furthermore, a motorized rotation stage with a half-wave plate (HWP) was used to rotate the polarization axis of the input laser beam.

**Multiple harmonics generation**

First, we shortly summarize the ways phase matching can be obtained within a β-BBO crystal and give the second and third order effective nonlinear coefficients for our experimental conditions. To describe the generation of harmonic up to 6$^{th}$ order, we abbreviate the generation process with the harmonic numbers like H2, H3, … H6 throughout this article, to avoid ambiguities; i.e., the term SHG could be used to denote both the second or the sixth harmonic.

The orientation of the β-BBO crystal and the used notations are indicated in the inset of Figure 1. Here, (x,y,z) are the laboratory coordinate axes. The laser beam propagates along the z-axis and its polarization angle ε is measured with respect to the x-axis. The axes (X,Y,Z) are set to the crystal with X∥a and Z∥c. The crystal originally is expected in the xy plane with x=X, y=Y and z=Z. It is cut with ϕ = 90°, which means that the propagation direction z is in the YZ plane. To reach phase matching, the crystal should be rotated around the X and y axes, which can be described by rotation matrices, R$_X$(θ) and R$_y$(ρ).

With β-BBO crystals, type I (oo→e) phase matching is possible for second harmonic generation meaning ε = 0° for the ordinary and ε = 90° for the extraordinary beams denoted with 'o' and 'e', respectively. The nonlinear interaction can be described by the reduced matrix formalism [30, 27, 28] for both H2 and H3 generations to obtain the second order d$_{eff}$ and third order C$_{eff}$ effective nonlinear coefficients. In the literature, inconsistencies can be found in the assignment of the crystal axes of the β-BBO crystal, for defining the susceptibility matrices and very different values for some matrix components are given [27, 28, 29, 31, 32]. However, one can recognize very small matrix components like $(d_{15}/d_{22})^2 = 3.3 \cdot 10^{-4}$ and $(C_{15}/C_{11})^2 = 2.1 \cdot 10^{-3}$, which represent the contributions of $d_{15}$ and $C_{15}$ to the generated signals. These values are smaller than the measurement errors and these matrix elements can be neglected under experimental conditions by setting $d_{15} \approx 0$ and $C_{15} \approx 0$. This essentially simplifies the handling of the β-BBO crystal, becoming optically equivalent to a hexagonal $D_{3h}$ crystal with point group 6m2, possessing higher symmetry than the original 3m point group. Furthermore, looking at the given measured values in [31] and considering measurement accuracies, we can approximate that $C_{11} \approx 3 \cdot C_{16}$ and $C_{33} \approx 2 \cdot C_{16}$. Using the mentioned simplification, we calculated the effective nonlinear coefficients. Experimentally, we can control the rotation ρ, so it is practical to explicitly retain ρ in the calculations. The generated H2 intensity is determined by the effective nonlinear coefficients

oo → e: $$d_{eff} \approx d_{22} \cos(\theta)[\cos^2(\rho) - \sin^2(\rho)\sin^2(\theta)] \tag{1a}$$

eo → e: $$d_{eff} \approx -d_{22} \cos^2(\theta) \sin(\theta) \sin(\rho). \tag{1b}$$

The case oo → e has a maximum $d_{eff} = d_{22}$ at ρ = 0° and θ = 0°. For phase matched H2, θ = 29.0° is required. The case eo → e has a maximum at ρ = 90° and θ = 35.3°. The ρ = 90° cannot be realized with a type I cut crystal, but choosing ρ ≠ 0° and the phase matching angle of θ = 41.9°, H2 can be generated, but lower efficiency is expected than in the oo → e case. The rotation with ρ affects θ and ϕ. Comparing the expressions for $d_{eff}$ (Eq. 1a and 1b) to the literature [27, 28], the values of θ and ϕ can be obtained from the following set of equations 2a and 2b:

$$\sin(3\phi) \approx \sin^2(\rho)\sin^2(\theta) - \cos^2(\rho) \tag{2a}$$

$$\cos(3\phi) \approx -\sin(\rho)\sin(\theta). \tag{2b}$$

The case of type I (ooo → e), H3 generation can be also realized using β-BBO crystals with effective nonlinear coefficient in Eq. (3a):

ooo → e: $$C_{eff} \approx C_{16} \sin^3(\rho) \cdot \cos^3(\theta) \cdot \sin(\theta) \tag{3a}$$

ooe → e: $$C_{eff} = C_{16} - C_{16} \sin^2(\rho) \cdot \sin^2(\theta) \cdot \cos^2(\theta) \tag{3b}$$

ooo → o: $$C_{eff} \approx 3 \cdot C_{16} - C_{16} \cdot \sin^4(\theta) \cdot \cos^4(\theta) \tag{3c}$$

eee → e: $$C_{eff} \approx 3 \cdot C_{16} - C_{16} \cdot \sin^4(\theta). \tag{3d}$$

Here, at ρ = 0°, $C_{eff}$ = 0 and it remains zero, independently of θ; H3 cannot be generated despite the fact that it can be phase matched at θ = 52.8°. The optimal choice would be ρ = 90° and θ = 30°, which cannot be realized with a H2 cut crystal directly. However, choosing a suitable polarization transforms the case to ooe → e, which can be phase matched at θ = 70.6° with $C_{eff}$ = $C_{16}$ (Eq. 3b) with ρ = 0° and H3 can be generated.

Considering cascaded nonlinear processes in addition to the direct generation of harmonics provides sometimes more phase matching possibilities, even when phase matching for the direct process is not possible. For cascaded processes, the sum of the individual k vectors $\Delta k_i$ should be zero [33] and it is not necessary that all participating processes are being separately phase matched:

$$\sum_i \Delta k_i = \frac{2\pi}{\lambda_l} \sum_i \Delta n_i = \frac{2\pi}{\lambda_l} \Delta n = 0. \tag{4}$$

Here $\lambda_l$ is the laser wavelength and $\Delta n$ represents the sum of the corresponding refractive indices of the contributing light beams, i. e., for oo → e second harmonic generation $\Delta n = n_o(\omega) + n_o(\omega) - 2n_e(\theta, 2\omega)$ and we use this more general definition. For cascaded processes, we need other $C_{eff}$ cases, which are given in Eqs. (3c) and (3d) for later use.

In the performed experiments, beyond the expected conversion of the Ti:sapphire laser pulses to their second harmonic, the appearance of further harmonics was observed. Although type II third harmonic generation is possible at $\theta = 70.6°$, this could not be realized under our experimental conditions. Therefore, we examined the case of type I H2 at $\theta = 29.0°$ and furthermore a possibility of cascaded H3 generation at $\theta = 44.6°$.

## *Results using type I phase matched second harmonic (from infrared up to H6)*

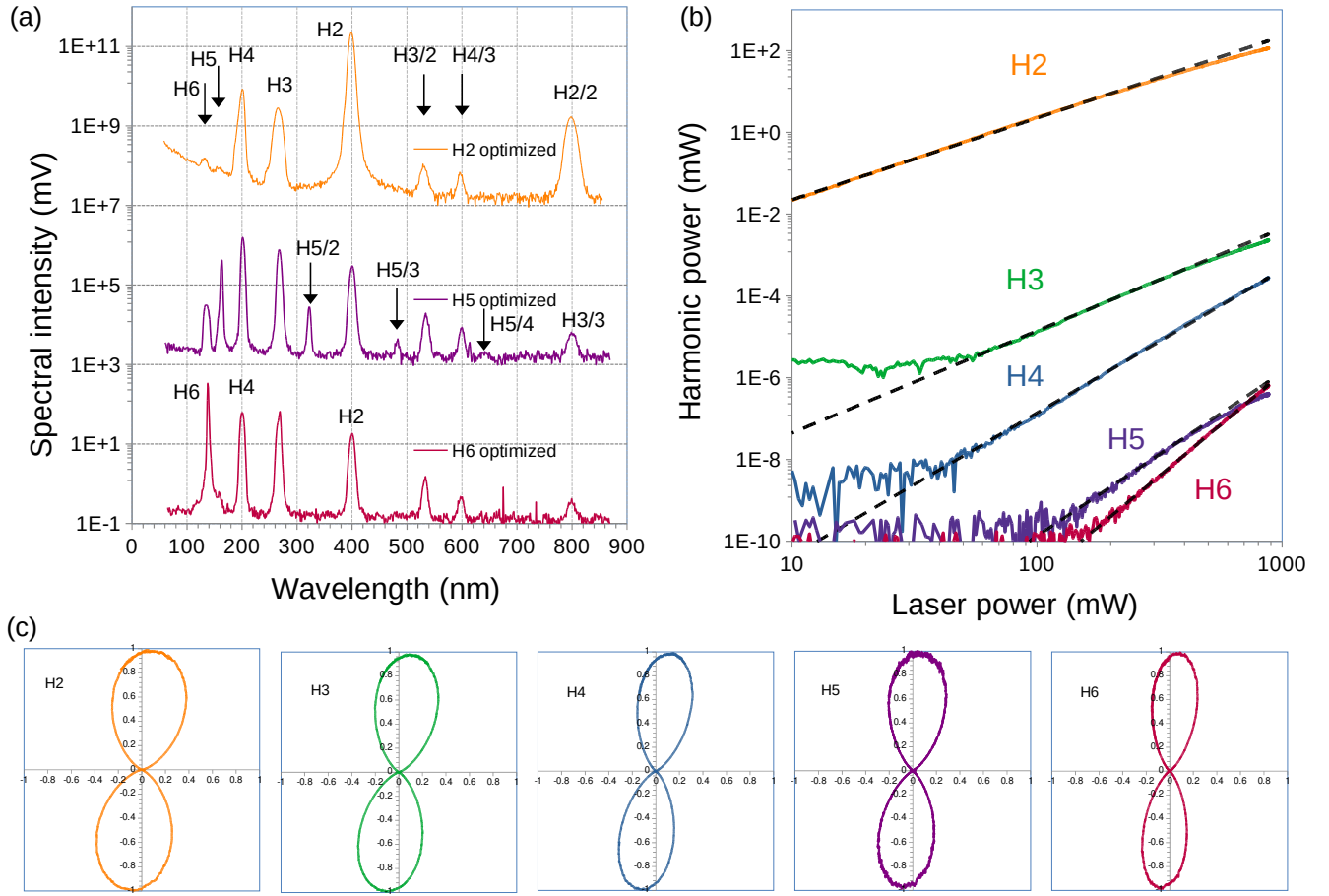

**Figure 2.** (a) Measured harmonic spectra at different positions of the VUV lens $f_3$ for selecting the harmonic line of interest. The spectra are shifted by factors of $10^4$ for better visibility. (b) Measured harmonic power dependence on the pump laser power and (c) on the polarization direction of the linearly polarized pump laser beam. The measurements were performed at $\theta = 29.0°$ and the vertical direction means ordinary laser beam. The dashed black lines in panel (b) represent the power law of the harmonic orders with ranks listed in Table 1.

In this chapter, we discuss the results obtained using the β-BBO crystal under type I phase matching with $\theta = 29.0°$ and $\phi = 90°$ when H2 at 400 nm is generated optimally. Beyond the second harmonic, further harmonics up to 6th order (133 nm) were observed. It was possible to selectively choose any of them by moving the VUV lens to focus the interesting harmonic beam into the input aperture of the spectrometer. Figure 2a depicts the spectra when the lens was set for H2, H5 or H6. All harmonic lines are best visible when the lens was set to H5 (160 nm). In this case, the second, third and fourth diffraction order of H5 (H5/2, H5/3, H5/4) at wavelengths of 320 nm, 480 nm and 640 nm can be also observed, additional to the usually observable ones, namely the second diffraction order of H3 (267 nm) at 533 nm, the third diffraction order of H4 (200 nm) at 600 nm and the second or third diffraction order of H2 or H3 at the position of the laser wavelength of 800 nm. When the lens focused H2 to the input aperture (curve in orange color), the presence of H5 and H6 is still weakly visible, but they are almost overwhelmed by the scattered background of the strong zero diffraction order of H2. By setting the VUV lens for H6 (magenta color), only a trace of H5 can be seen in the pedestal of H6 at 160 nm and a weak H7 can be also recognized at 114 nm, being at the edge of our observable wavelength range.

Higher than second order harmonics can be generated directly or by cascaded processes. In a previous work on β-BBO nano-crystals [26], the measured polarization dependence of the different harmonic signals suggested a direct generation by high order nonlinear processes. In our case, to determine/identify the processes contributing to the multiple harmonic generation, the dependence of the generated harmonic power on the polarization direction of the laser beam was measured in a similar way and depicted in Figure 2c. We used a half-wave-plate (HWP) at the input laser beam to rotate the polarization direction of the linearly polarized laser beam. The estimated laser intensity in the focus was ≈ 200 GW/cm², therefore the generation process can be handled as perturbative, which is in agreement with the observation that the H2 signal can be described by a $\cos^4(\varepsilon+\delta)$ function, where $\varepsilon$ is the polarization direction of the laser beam as given in the inset of Figure 1. The small $\delta = 6°$ rotation is the consequence of a rotation of the crystal around the y-axis, which shifts the maximum of $d_{eff}$ somewhat away from $\varepsilon = 0$. All other polarization curves follow H2 and show very similar $\cos^{2r}(\varepsilon+\delta)$ behavior with r > 2, suggesting that the H2 beam plays a dominant role in their generation. It is further confirmed by rotating the crystal around the X-axis and detuning from the H2 phase matching angle, which also extinguished the higher order harmonics together with H2. We thus conclude that higher order harmonics are not generated directly by high order $\chi^{(q)}$ nonlinear interactions, but by cascades of low order nonlinear interactions.

| q | 1 | 2 | 3 | 4 | 5 | 6 |
|---|---|---|---|---|---|---|
| λ (nm) | 800 | 400 | 266 | 200 | 160 | 133 |
| P (29°) | 880 mW* | 250 mW | 2-5 µW | 0.3-0.6 µW | 0.8-1.5 nW | 1-2 nW |
| r (29°) |  | 2 | 2.5 | 3.5 | 4 | 5 |
| P (44.6°) | 880 mW* | 15 mW | 50-100 µW | < 0.1 nW |  |  |
| r (44.6°) |  | 2.3 | 3 |  |  |  |

**Table 1.** The measured harmonic powers and the rank of the nonlinear processes in both cases of type I H2 alignment at θ = 29.0° and cascaded H3 at θ = 44.6°. *The laser power is given at the input, before interaction. Note that the power of the H5 and H6 VUV harmonics are clearly sufficient for several spectroscopic applications.

To investigate this behavior further, we repeated the polarization measurement by placing a polarizer into the laser beam after the HWP. It would ensure that the polarization is always set to the ordinary direction and the laser intensity changes proportionally to cos²ε by the HWP rotation. Because a film polarizer became damaged at the higher laser powers, we used a 0.1 mm thin Brewster plate, which transmitted the ordinary and essentially suppressed the extraordinary beam component. From the depicted results in Figure 2b, two important conclusions can be drawn. First, a weak effect of saturation can be seen for harmonic orders of H2, H3 and H5, but H4 and H6 show no sign of saturation within this intensity range. Secondly, before saturation, the harmonic intensities well follow the power law of $I_q \sim I^r$ but sometimes the r rank of the nonlinear process r ≠ q. We summarized the experimental results in Table 1.

To determine phase matching angles, different sets of Sellmeier equations can be chosen. Unfortunately, they mainly concentrate on the infrared spectral range. We apply the Sellmeier equations of ref. [34], because it includes measurement points in the ultraviolet spectral range down to 186 nm and can be presumably more accurate for our interest. A type I (oo→e) phase matching based on the ω+ω→2ω process at θ = 29.0° is predicted. Direct type II H3 (ω+ω+ω→3ω, ooe→e) can also be generated, however at a very different angle of θ = 70.6°, being too far to be intense at 29°. We looked for cascades, where the first $\chi^{(2)}$ process is of type I H2 (ω+ω→2ω, oo→e) followed by (an)other $\chi^{(2)}$ or $\chi^{(3)}$ process(es). The possible cascades are presented in Figure 3.

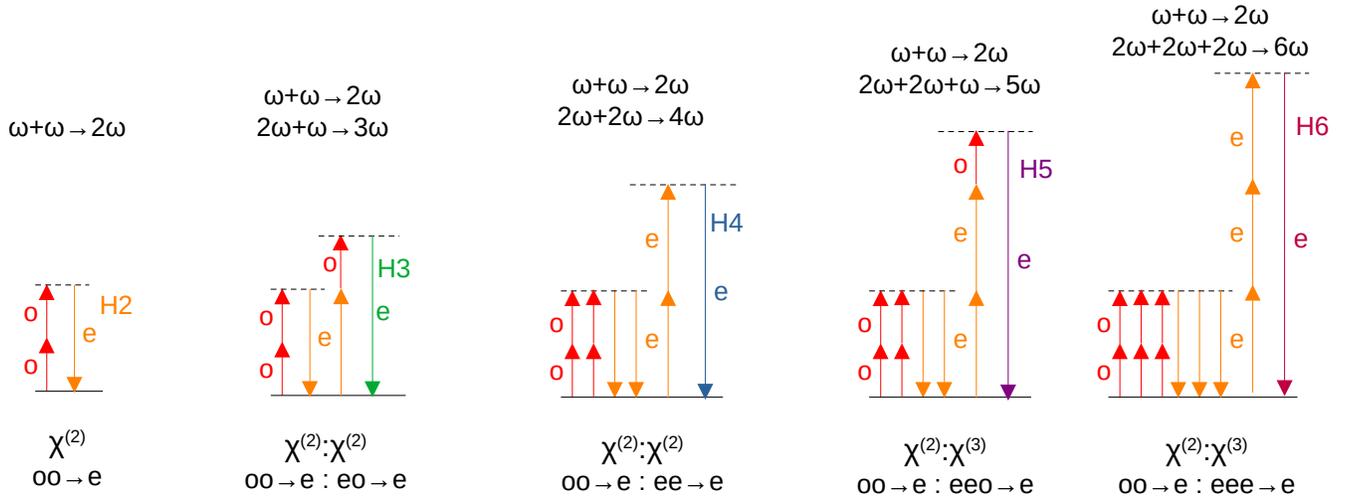

**Figure 3:** Representation of the generation of the different harmonics. All of them start with a type I (oo→e) H2 at θ = 29° and are followed by a second or third order cascade. The length of the arrows is proportional to the photon energy, and 'o' and 'e' means ordinary and extraordinary beam. The color of the arrows also indicates the harmonic order, namely laser: red (↑), H2: orange (↑), H3: green (↑), H4: blue (↑), H5: violet (↑), H6: magenta (↑).

In the case of H3, the generation is possible due to a $\chi^{(2)}:\chi^{(2)}$ cascade with 2ω+ω→3ω, eo→e, or a $\chi^{(2)}:\chi^{(3)}$ cascade with 2ω+2ω-ω→3ω, eeo→e, as second steps (Fig. 3). According to Eq. (4), the phase matching condition for both cases is $\Delta n = 3n_o(\omega) - 3n_e(3\omega) = 0$, gives θ = 52.8°. This is not phase matched at θ = 29°, but still can produce weaker radiation by a higher order maximum of the phase matching factor

$$F_{PM}=\frac{\sin^2\left(\frac{1}{2}\Delta kL\right)}{\left(\frac{1}{2}\Delta kL\right)^2}, \qquad (5)$$

when

$$\frac{1}{2}\Delta kL=\frac{\pi}{2}+m\pi \qquad (6)$$

and then the harmonic intensity is proportional to

$$I_H \propto \frac{1}{\left(\frac{1}{2}\Delta kL\right)^2}=\frac{\lambda_l^2}{(\pi\Delta nL)^2}=\frac{1}{\left(\frac{\pi}{2}+m\pi\right)^2}. \qquad (7)$$

For the above-mentioned case of H3, m ≈ 21, gives a relative intensity of ≈$10^{-4}$, which is feasible looking to the low H3 power in Table 1. The contribution of the $\chi^{(2)}$:$\chi^{(3)}$ cascade is probably weaker than the $\chi^{(2)}$:$\chi^{(2)}$ cascade, therefore we did not indicate it in Figure 3. The case of H4 is very similar to H3, the generation can happen via a $\chi^{(2)}$:$\chi^{(2)}$ cascade, but using only the H2 beam in the second step like 2ω+2ω→4ω, ee→e. The phase matching condition $\Delta n=4n_o(\omega)-4n_e(4\omega)=0$ cannot be fulfilled. It yields m ≈ 89 with a relative intensity of ≈$10^{-5}$ which is in good agreement with H4 being one order of magnitude weaker, compared to H3 in Table 1. The cases of H5 and H6 are very different, because the crystal becomes absorbing and the phase matching factor can be written as [11]

$$F_{PM}=\frac{e^{-\alpha L}-2e^{-\alpha L/2}\cos(\Delta kL)+1}{(\Delta kL)^2+(\alpha L/2)^2}, \qquad (5)$$

it is dependent on the absorption coefficient α, which can be expressed by the absorption length $\alpha=L_a^{-1}$. In this case, the harmonic generation is absorption limited and only the last ≈$L_a$ length participates effectively to the harmonic signal, L ≈ $L_a$. Unfortunately, the complex refractive index of the β-BBO crystal in the VUV range has not been measured yet, but $L_a$ should be in the 10-100 nm range, comparing to materials with similar band gaps of ≈ 6 eV. Such a short $L_a$ length essentially broadens the angle acceptance range of the phase matching and the process can be considered phase matched at any θ. However, the short interaction length dramatically decreases the output signal, $I_H \propto (L_a/L)^2$, yielding a reduction factor of ≈$10^{-6}$-$10^{-8}$, which is in good agreement with the measured power given in Table 1. The phase mismatch, similarly to H3 and H4, is $\Delta n=5n_o(\omega)-5n_e(5\omega)$ and $\Delta n=6n_o(\omega)-6n_e(6\omega)$ for the $\chi^{(2)}$:$\chi^{(3)}$ cascades, namely 2ω+2ω+ω→5ω, eeo→e, and 2ω+2ω+2ω→6ω, eee→e, respectively. H5 could have been also generated via $\chi^{(2)}$:$\chi^{(2)}$ with 3ω+2ω→5ω, ee→e, but this contribution should be weak because H3 is much weaker than H2.

To shortly summarize our experimental results on cascaded processes:

a) In the cases of H4 and H6, no saturation of the signals is observed, because only H2 takes part in their generation. In contrast, in the generation of H2, H3 and H5, the pump laser beam also contributes and is essentially depleted by H2 generation, resulting in a saturation.

b) At the measured alignment of θ = 29.0°, only H2 generation is phase matched and produces a strong signal. H3 and H4 cannot and are not phase matched simultaneously, and produce orders of magnitude weaker signals. H5 and H6 can be considered phase matched, but because of the very short production length, the generated harmonic powers are even smaller. We note that while the power of H5 appears suppressed in comparison, it is still comparable or even higher to the power reached by higher non-perturbative intensities of ~ 1000 GW/cm² from AlN film [3, 35] with a similar band gap. Additionally, H6 also appeared with similar power to H5 as the consequence of the strong H2 beam.

*Results for phase matched cascaded third harmonic*

As mentioned earlier, not only type I but also type II (oe→e) phase matching of H2 can be realized at θ= 41.9°. Interestingly, it is close to a geometry to generate phase matched H3 by cascaded $\chi^{(2)}{:}\chi^{(3)}$ at θ= 44.6°. These possible cascades are shown in Figure 4a. In the first step, H2 is generated via oo→e or oo→o and in the second step the generated H2 e- or o-polarized photons together with e-polarized pump laser photons generate H3, via eee→e or ooe→e, 2ω+2ω-ω→3ω. An alternative path is shown in Figure 4(b), where H3 is generated by a $\chi^{(3)}{:}\chi^{(3)}$ cascade. Here, in the first step, weak, non-phase matched H3 is generated via ooo→o and in the second step, these o-polarized H3 photons, interacting with o- and e- polarized pump laser photons, generate a phase matched H3 e-polarized beam via ooe→e, 3ω+ω-ω→3ω. For each involved $\chi^{(3)}$ process, $C_{eff}$ is large as can be seen in Eqs. (3a)-(3d). Remarkably, none of the mentioned intermediate steps are phase matched at θ = 44.6°, but the cascade using four o-polarized and one e-polarized laser photons is phase matched with $\Delta n = 4n_o(\omega) - n_e(\omega) - 3n_e(3\omega) = 0$. Independently, type II H2 is generated close to its phase matching angle and within its angle acceptance interval with somewhat reduced power. The measured spectra are presented in Figure 5(a). When the spectrometer is set to optimally detect H3 (green curve) a weak H2 peak can also be seen. As it was seen in the previous chapter, H2 should generate H4 by 2ω+2ω→4ω, which can be observed if the spectrograph is set to optimally measure H4, as shown by the blue curve. For this measurement, higher gain (driving voltage) for the photomultiplier was used and the H3 saturated the detector. While H3 is strong, 50-100 µW, H2 is weak ~15 mW and consequently H4 is very weak, below 0.1 nW. The values are listed also in Table 1 for comparison. It was not possible to observe H5 and H6 because both of them require strong H2. The absence of H5 and H6 further proves that they are not generated by direct $\chi^{(5)}$ or $\chi^{(6)}$ processes, although a strong pump laser beam is present.

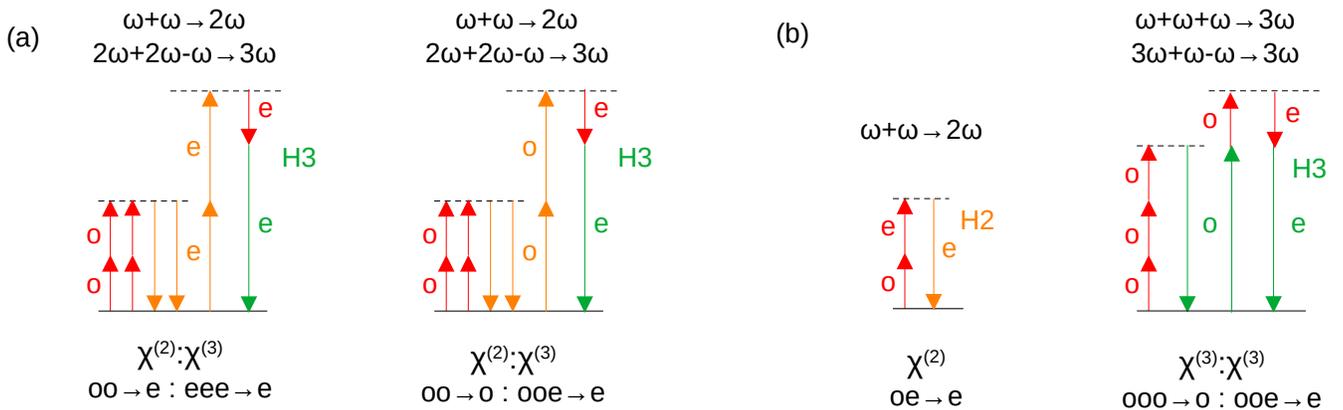

**Figure 4:** Phase matched H3 by cascaded harmonic generation at θ = 44.6°. (a) and (b) show the corresponding $\chi^{(2)}{:}\chi^{(3)}$ and $\chi^{(3)}{:}\chi^{(3)}$ cascades. Panel (b) shows also the independent type II (oe→e) H2 generation "phase matchable" at θ = 41.9°. As H3 is within the H2 generation angle acceptance range, both harmonic orders are generated.

Furthermore, we measured the dependence of the harmonic signals on the polarization direction of the pump laser beam, as depicted in Figure 5(c). In the case of type II H2, the generated harmonic field is determined by both the o- and e-polarized laser beam and consequently $E(2\omega) \propto E^2(\omega)\cos(\epsilon)\sin(\epsilon)$ with a maximal signal at ε = 45°, which nicely can be seen in Figure 5(c). In the case of H4, the polarization curves clearly indicate that H4 is generated by H2 and some signal from the scattered strong H3 (dotted line) is added. A small rotation of the curve is a consequence of the used small (ρ = 20°) rotation of the crystal around the y-axis to obtain the optimal H3 signal. The curve of the H3 is more rotated as the H3 in Figure 2, because beyond the four o-polarized laser photons, one e-polarized also takes part in the generation and $E(3\omega) \propto E^4(\omega)\overline{E}(\omega)\cos^4(\epsilon)\sin(\epsilon)$. Using these assumptions for the polarization dependence, we measured the intensity dependent harmonic signals (Figure 5(b)), similarly to the previous chapter, by using a thin plate in Brewster angle at the input laser beam. The calculated (dashed black) lines nicely follow the measured ones (solid green and orange lines).

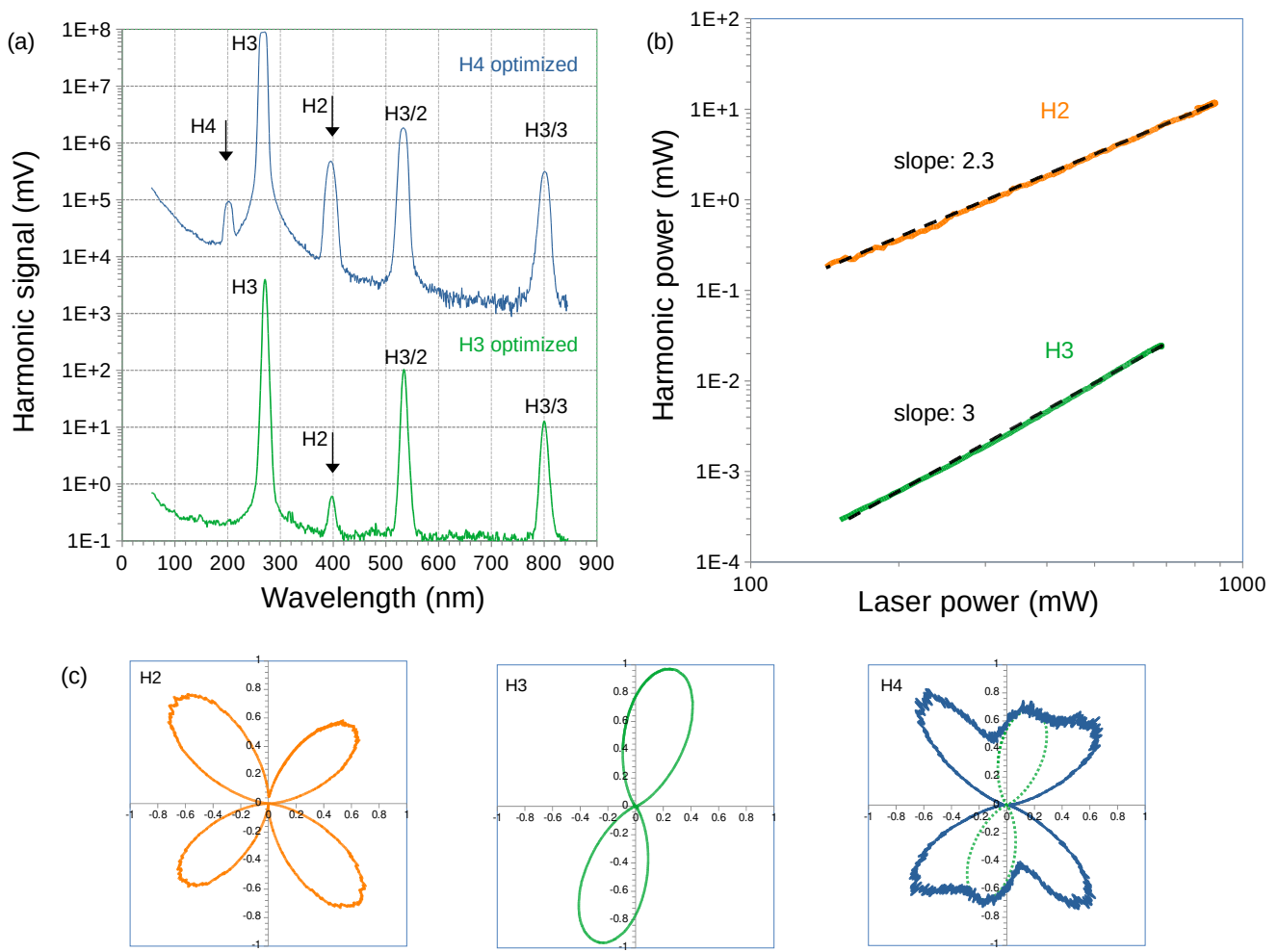

**Figure 5.** (a) Measured harmonic spectra in the case of type II H2 and $\chi^{(2)}$:$\chi^{(3)}$ phase matched H3 at θ = 44.6°. The spectrum in blue color is shifted (multiplied with $10^4$) for better visibility. (b) Pump laser power dependence of the yields of H2 and H3. (c) Measured polarization dependence of the harmonic signals.

**Summary**

We demonstrated the generation of multiple harmonics in a β-BBO crystal containing both even and odd harmonics. When the crystal was aligned for type I second harmonic generation, the strong second harmonic beam together with the fundamental laser beam acted as a two-color pump and produced harmonics up to the 6$^{th}$ order, extending into the vacuum ultraviolet and reaching 133 nm. We identified the generation process as cascades of low order $\chi^{(2)}$ and $\chi^{(3)}$ nonlinear interactions and both $\chi^{(2)}{:}\chi^{(2)}$ and $\chi^{(2)}{:}\chi^{(3)}$ cascades were observed. The dominance of the cascades over direct high order interactions were proven by polarization measurements. Additionally, we realized cascaded third and fourth harmonic generation near the type II phase matched condition, demonstrating that the fourth harmonic was directly generated by the second harmonic beam and the third harmonic was independent.

In the context of practical applications, we can consider the advantages of the developed source. The multi-harmonic source was directly driven by a Ti:sapphire oscillator, operated in the 100 MHz frequency range, which makes it attractive as a frequency comb for high precision measurements in the ultraviolet and vacuum ultraviolet spectral range. Although, more harmonics cannot be phase matched simultaneously and for the 5$^{th}$ and 6$^{th}$ harmonics the β-BBO crystal is absorbing, limiting the harmonic signals, the generated harmonics are still intense and even the weakest ones are 3 orders of magnitude above the measurement noise limit, which makes the source suitable for many spectroscopic applications.

It was possible to generate relatively strong H5 and H6 harmonics in the VUV spectral range, where the dispersion of β-BBO and also the transmission of the used protecting coating of the crystal is not known. The presented results might encourage the examination of the crystal and the development of a highly transparent coating in the VUV range.

**Funding**

This work is part of the ThoriumNuclearClock project that has been funded by the European Research Council (ERC) under the European Union's Horizon 2020 research and innovation program (Grant Agreement No. 856415) and the Austrian Science Fund (FWF) [Grant DOI: 10.55776/F1004, 10.55776/J4834, 10.55776/PIN9526523]. This work has received funding from the European Partnership on Metrology, co-financed by the European Union's Horizon Europe Research and Innovation Programme and by the Participating States, under grant number 23FUN03 HIOC. We acknowledge support from the Österreichische Nationalstiftung für Forschung, Technologie und Entwicklung (AQUnet project).